\documentclass[aps,prx,reprint,superscriptaddress,amsmath]{revtex4-2}
\usepackage{physics}
\usepackage{graphicx}
\usepackage{gensymb}

\begin{document}


\title{Effects of Floquet Engineering on the Coherent Exciton Dynamics in Monolayer WS$_2$}

\author{Mitchell A.\ Conway}
\author{Stuart K.\ Earl}
\author{Jack B.\ Muir}
\affiliation{Optical Sciences Centre, Swinburne University of Technology, Hawthorn, 3122, Victoria, Australia}
\affiliation{ARC Centre of Excellence in Future Low-Energy Electronics Technologies, Swinburne University of Technology, Hawthorn, 3122, Victoria, Australia}

\author{Thi-Hai-Yen Vu}
\affiliation{\small \textit{ARC Centre of Excellence in Future Low-Energy Electronics Technology, Monash University, Clayton, 3800, Victoria, Australia}}
\affiliation{\small \textit{School of Physics and Astronomy, Monash University, Clayton, 3800, Victoria, Australia}}

\author{Jonathan O.\ Tollerud}
\affiliation{Optical Sciences Centre, Swinburne University of Technology, Hawthorn, 3122, Victoria, Australia}
\affiliation{ARC Centre of Excellence in Future Low-Energy Electronics Technologies, Swinburne University of Technology, Hawthorn, 3122, Victoria, Australia}

\author{Kenji Watanabe}
\affiliation{Research Center for Functional Materials, National Institute for Materials Science, Tsukuba, Ibaraki 305-044, Japan}

\author{Takashi Taniguchi}
\affiliation{International Center for Materials Nanoarchitectonics, National Institute for Materials Science, Tsukuba, Ibaraki 305-0044, Japan}

\author{Michael S.\ Fuhrer}
\affiliation{\small \textit{ARC Centre of Excellence in Future Low-Energy Electronics Technology, Monash University, Clayton, 3800, Victoria, Australia}}
\affiliation{\small \textit{School of Physics and Astronomy, Monash University, Clayton, 3800, Victoria, Australia}}

\author{Mark T.\ Edmonds}
\affiliation{\small \textit{ARC Centre of Excellence in Future Low-Energy Electronics Technology, Monash University, Clayton, 3800, Victoria, Australia}}
\affiliation{\small \textit{School of Physics and Astronomy, Monash University, Clayton, 3800, Victoria, Australia}}
\affiliation{\small \textit{ANFF-VIC Technology Fellow, Melbourne Centre for Nanofabrication, Victorian Node of the Australian National Fabrication Facility, Clayton, VIC 3168, Australia}}

\author{Jeffrey A.\ Davis}
\affiliation{Optical Sciences Centre, Swinburne University of Technology, Hawthorn, 3122, Victoria, Australia}
\affiliation{ARC Centre of Excellence in Future Low-Energy Electronics Technologies, Swinburne University of Technology, Hawthorn, 3122, Victoria, Australia}
\email[]{jdavis@swin.edu.au}


\begin{abstract}
Coherent optical manipulation of electronic bandstructures via Floquet Engineering is a promising means to control quantum systems on an ultrafast timescale. However, the ultrafast switching on/off of the driving field comes with questions regarding the limits of validity of the Floquet formalism, which is defined for an infinite periodic drive, and to what extent the transient changes can be driven adibatically.
Experimentally addressing these questions has been difficult, in large part due to the absence of an established technique to measure coherent dynamics through the duration of the pulse. Here, using multidimensional coherent spectroscopy we explicitly excite, control, and probe a coherent superposition of excitons in the $K$ and $K^\prime$ valleys in monolayer WS$_2$. With a circularly polarized, red-detuned, pump pulse, the degeneracy of the $K$ and $K^\prime$ excitons can be lifted and the phase of the coherence rotated. We demonstrate phase rotations during the 100 fs driving pulse that exceed $\pi$, and show that this can be described by a combination of the AC-Stark shift of excitons in one valley and  Bloch-Siegert shift of excitons in the opposite valley. Despite showing a smooth evolution of the phase that directly follows the intensity envelope of the pump pulse, the process is not perfectly adiabatic. By measuring the magnitude of the macroscopic coherence as it evolves before, during, and after the pump pulse we show that there is additional decoherence caused by power broadening in the presence of the pump. This non-adiabaticity may be a problem for many applications, such as manipulating q-bits in quantum information processing, however these measurements also suggest ways such effects can be minimised or eliminated.
\end{abstract}

\maketitle

\section{\label{sec:I}Introduction}

Using light to manipulate the properties of condensed matter can enable ultrafast control of charge carrier densities \cite{matsubara2015ultrafast}, spin \cite{bassett2014ultrafast,kampfrath2011coherent} and pseudospin \cite{ye2017optical} of quasiparticles, band structures \cite{Wang2013} or quantum \cite{Hellmann2010,Iwai,Fausti2011} and topological \cite{mciver2020light} phase transitions. These transient changes can be used for ultrafast switching in classical information processing, implemented in parallel and with spatial control of the light to enable neuromorphic computing \cite{ballarini2020polaritonic}, or to enable future quantum technologies. The extent to which this ultrafast control can proceed adiabatically will determine the feasibility of realizing quantum technologies driven by ultrafast laser pulses. In many cases, however, experimentally determining the adiabaticity of these ultrafast processes is difficult, with no established method to measure the coherent electron dynamics through the pulse duration \cite{earl2021coherent,uchida2022diabatic}.

Semiconducting monolayer transition metal dichalcogenides (TMDCs) offer unique opportunities for control \cite{earl2021coherent,kim2014ultrafast,sie2015valley,sie2016observation,sie2017blochsiegert,ye2017optical,lamountain2018valley,yong2018biexcitonic,cunningham2019resonant,morrow2020quantum,slobodeniuk2022semiconductor,slobodeniuk2022giant,kobayashi2023floquet}. In these materials, the degenerate, direct band gaps at the $K$ and $K^\prime$ points in momentum space couple selectively to left- and right-circularly polarized light, respectively \cite{xiao2012coupled,mak2014valley}. This facilitates valley selective control of the bands, as has been demonstrated through the valley selective AC-Stark effect \cite{kim2014ultrafast,sie2015valley}, and Bloch-Siegert effect \cite{sie2017blochsiegert}. These effects can be viewed as specific cases of Floquet engineering, where the periodic electric field of the light dresses the equilibrium bandstructure. Control of bandstructures via Floquet engineering is attracting significant interest due to the potential to realise ultrafast transitions between different quantum and topological phases \cite{Oka_review,Schuler2022}. For example, the opposite Berry curvature at each valley in monolayer TMDCs has subsequently led to proposals to realise a Floquet topological insulator phase \cite{sie2015valley,claassen2016all}, similar to what has been achieved in graphene \cite{mciver2020light}. The utility of such phase transitions is dependent on the process proceeding adiabatically, which in part requires that the switching on of the Floquet driving does not introduce any non-adiabatic dynamics. 

The valley selective AC-Stark effect has also been used to manipulate the phase of an inter-valley coherence in WSe$_2$ \cite{ye2017optical}. That is, a coherent superposition of degenerate $K$ and $K^\prime$ excitons (${\ket{X_K}\bra{X_{K^\prime}}}$) was generated \cite{jones2013optical,hao2016direct} and its phase rotated by the valley selective AC-Stark effect lifting the degeneracy for the duration of the driving pulse. Coherent optical manipulation such as this has potential for quantum information processing, but also provides a means to determine whether the turning on/off of the Floquet drive and shifting of the bands is adiabatic. Signatures that it is would include a smooth (and reversible) evolution of the phase and no additional decoherence.


Here we utilise multidimensional coherent spectroscopy (MDCS) \cite{moody2015intrinsic,hao2016coherent,conway2022direct,muir2022interactions,tollerud2017coherent} to create an inter-valley coherence in monolayer WS$_2$ and measure the pump induced changes to the amplitude and phase. While previous measurements of the phase rotation of the inter-valley coherence relied on measuring the polarization of the emitted photoluminescence and suffered from temporal and spatial averaging effects, our approach directly measures the phase and amplitude of the coherence within the $30 \pm 2$ fs window of the final MDCS pulse. We demonstrate a $\pi$ phase shift of the inter-valley coherence that arises primarily from the AC-Stark shift in one valley, but is offset by the Bloch-Siegert shift in the opposing valley. With 10 fs time resolution, these measurements show that the phase evolves smoothly and that the shift of the bands closely follows the instantaneous intensity of the pump pulse. However, by monitoring the amplitude of the third order signal before, during, and after the pump pulse we reveal that the total macroscopic coherence is reduced irreversibly as a result of power broadening of the exciton states, thereby indicating that this band structure control is not purely adiabatic. 
 
%


\begin{figure}
    \includegraphics{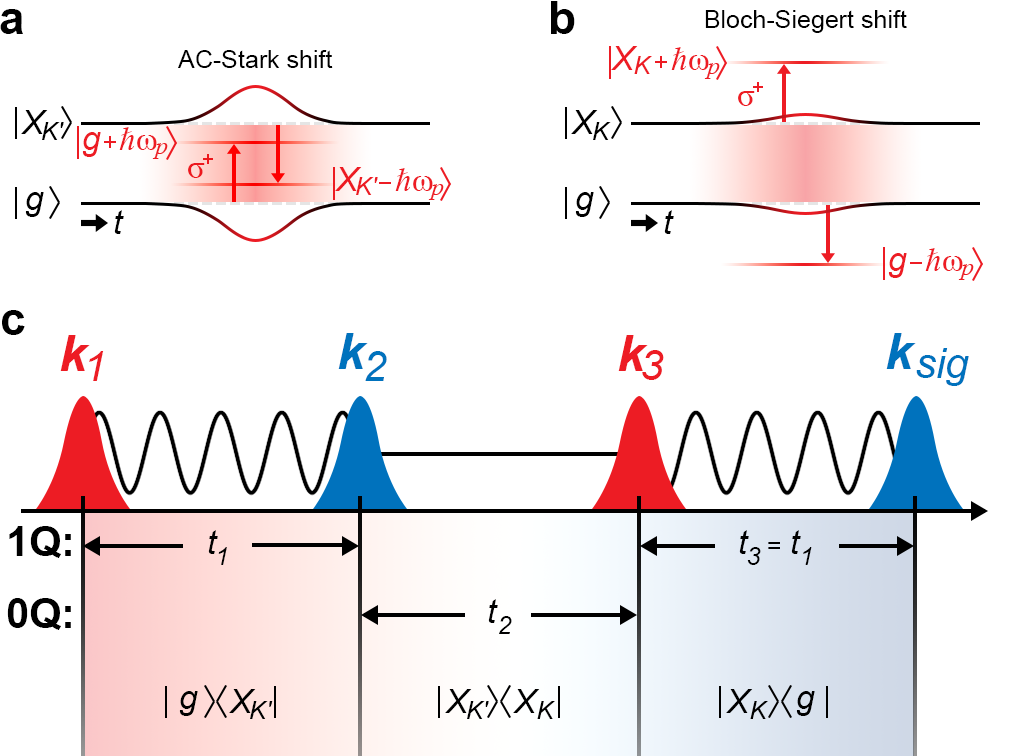}
    \caption{\label{fig:1}(a) Energy diagram of the $K^\prime$ valley. The energies of the $K^\prime$ valley exciton state $\ket{X_{K^\prime}}$ and ground state $\ket{g}$ are unperturbed until the arrival of the pump. In the presence of the pump, the coherent light-matter interaction between the equilibrium bands and a red-detuned $\sigma^+$ polarized pump, with energy $\hbar\omega_p$, produces Floquet-bands $\ket{g+\hbar\omega_p}$ and $\ket{X_{K^\prime}-\hbar\omega_p}$, which interact repulsively with $\ket{X_{K^\prime}}$ and $\ket{g}$, respectively, inducing an AC-Stark shift. The magnitude of the AC-Stark shift is expected to follow the instantaneous intensity of the pump pulse\cite{earl2021coherent} (red lines), and the states return to their unpumped energies at the end of the finite pump duration. (b) In the $K$ valley, the red-detuned $\sigma^+$ polarized pump causes a Bloch-Siegert shift as the equilibrium bands repel from the $\ket{g-\hbar\omega_p}$ and $\ket{X_{K}+\hbar\omega_p}$ Floquet-bands. (c) The rephasing pulse ordering used, where either the $t_1$ or $t_2$ delay is scanned to acquire a 1Q or 0Q signal, respectively. In our measurements, the MDCS beams alternate helicity, represented by the colors of each pulse, where red corresponds to $\sigma^+$, and blue to $\sigma^-$ circularly polarized pulses).}
\end{figure}

\section{\label{sec:II}Experimental methods}

In these MDCS measurements the first two optical excitation pulses (with wavevector $\textbf{\textit{k}}_1$ and $\textbf{\textit{k}}_2$ and opposite circular polarization) resonantly excite the inter-valley exciton coherence ${\ket{X_K}\bra{X_{K^\prime}}}$, as shown schematically in Fig.\ \ref{fig:1}(c). This zero-quantum (0Q) coherence (so-called because the energy separation of the states in superposition is small relative to the laser photon energy) evolves during the time delay $t_2$ until the third pulse (with wavevector $\textbf{\textit{k}}_3$) converts the 0Q-coherence into a third-order polarization (${\ket{X_K}\bra{g}}$), which radiates as the four-wave mixing signal \cite{tollerud2017coherent,moody2017advances}. Using heterodyne detection, we measure both the amplitude and phase of this signal as a function of emission energy ($\hbar\omega_3$) and inter-pulse delays \cite{tollerud2017coherent}, as described in Appendix \ref{appen_MDCS}. 
In these ensemble measurements, the signal amplitude reflects the degree of coherence and its evolution is used to reveal and quantify decoherence. The signal phase is determined by the phase of each excitation pulse and the phase evolution of coherences during the inter-pulse delays: $\phi_{sig} = -\phi_1 + \phi_2 + \phi_3 + \omega_1 t_1 + \omega_2 t_2 + \omega_3 t_3$. Here, $\hbar\omega_n$ is equal to the energy separation of states in superposition during the $t_{n}$ inter-pulse delay. To study the dynamics of the inter-valley exciton coherence, we scan $t_2$. Under equilibrium conditions, the degeneracy of the $K$ and $K^\prime$ excitons means there is no evolution of the phase of the inter-valley coherence (and hence no evolution in the phase of the signal) as a function of $t_2$. Consistent with previous measurements \cite{hao2016direct,hao2016coherent,muir2022interactions,jakubczyk2016radiatively}, we show that this inter-valley coherence has a decoherence time of $210 \pm 10$ fs. 

The introduction of the red-detuned $\sigma^+$ polarized pump pulse 150 fs after $\textbf{\textit{k}}_1$ and $\textbf{\textit{k}}_2$ lifts the degeneracy of the $K$ and $K^\prime$ excitons, as depicted in Fig.\ \ref{fig:1}(a) and (b). As a result, the phase of the inter-valley coherence evolves while the pump is present, which is reflected in a phase rotation of the measured signal.

With a $\sigma^+$ polarized pump pulse, the AC-Stark effect and Bloch-Siegert effect cause a blue-shift of the $K^\prime$ and $K$ exciton state energies, respectively \cite{sie2017blochsiegert}, as depicted in Fig.\ \ref{fig:1}(a) and (b). 
While the shift in both valleys is in the same direction the magnitudes of the shift are different, which breaks the K-K' degeneracy, albeit with a reduced energy difference relative to an isolated AC-Stark shift. To isolate the effects of the AC-Stark and Bloch-Siegert effects, we measure the signal evolution as a function of $t_1$ without any optical pump and with a circularly polarized pump arriving 230 fs before $\textbf{\textit{k}}_2$. In this case, the $\sigma^+$ polarized $\textbf{\textit{k}}_1$ pulse generates a coherence between the ground state, $g$, and the $K^\prime$ valley exciton state: ${\ket{g}\bra{X_{K^\prime}}}$, which evolves over $t_1$ with a frequency given by the A-exciton energy in monolayer WS$_2$ of 2.07 eV. With the $\sigma^+$ ($\sigma^-$) polarized pump arriving 230 fs before $\textbf{\textit{k}}_2$, the AC-Stark effect (Bloch-Siegert effect) will shift the $K^\prime$ valley exciton and perturb the evolution of the coherence.

\section{\label{sec:III}Results and discussion}
\subsection{\label{sec:A}Manipulation of coherence phase}

\begin{figure}
    \includegraphics{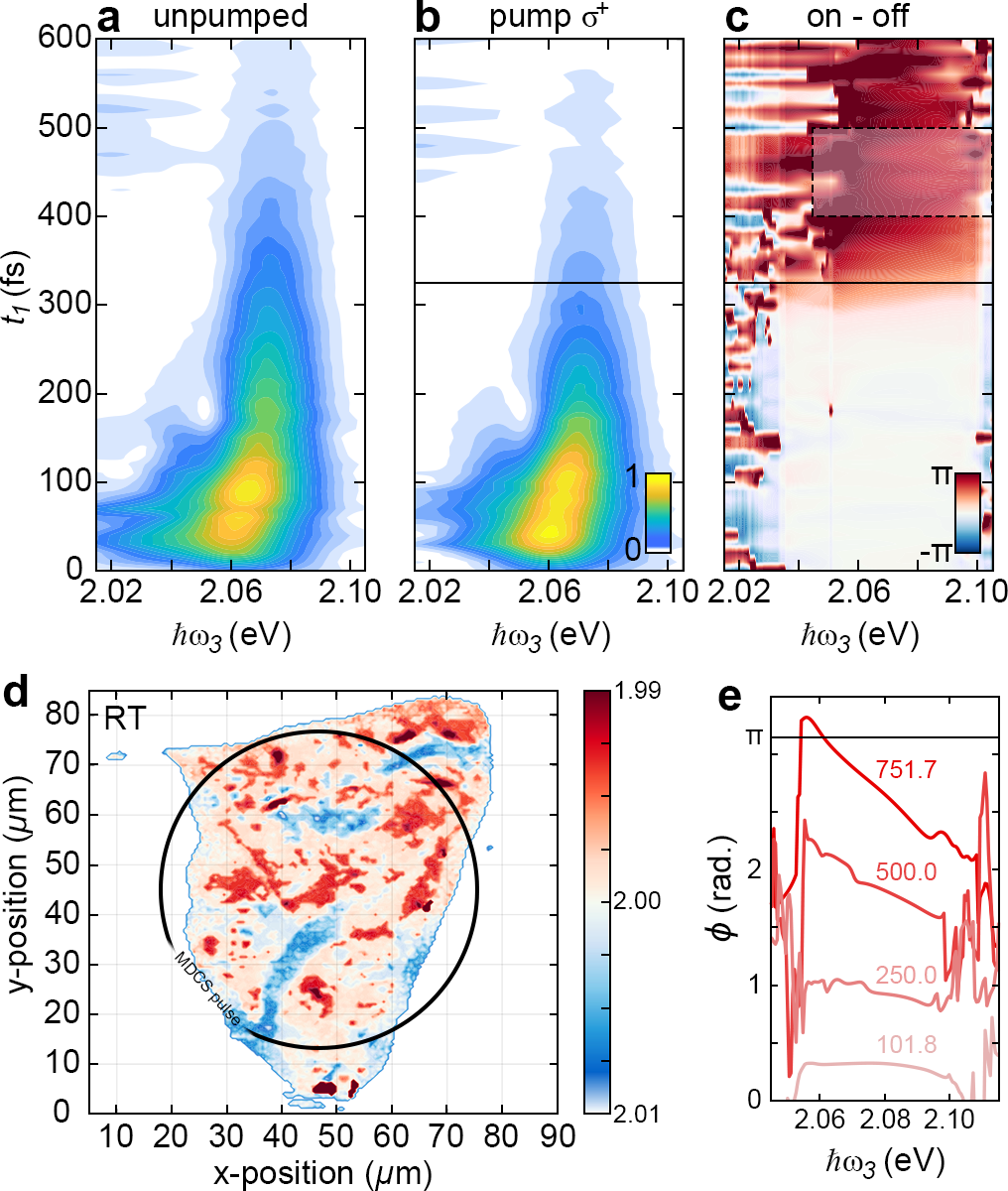}
    \caption{\label{fig:2}(a) Signal amplitude as a function of $t_1$ and $\hbar\omega_3$ of monolayer WS$_2$ at 4 K, without the pump. (b) Signal amplitude with a $\sigma^+$-polarized pump centered at 325 fs, normalised to the peak of the pump-off signal. There is a clear drop in the amplitude of the signal for $t_1>250$ fs, suggesting some additional loss of coherence as a result of the pump being present. (c) The difference between the phase of the signal with and without the pump (i.e.\ $\phi_{on}-\phi_{off}$) as a function of $t_1$ and $\hbar\omega_3$. The pump produces an AC-Stark shift of the $K^\prime$ valley exciton, which is responsible for inducing the phase rotation. (d) Spatially resolved PL of our monolayer WS$_2$ flake, where the colors represent the exciton peak energy in eV, reveals that the exciton energy varies significantly across the laser spots used for the MDCS measurements. The variations arise due to micro-cracks, bubbling hBN encapsulation, and other defects. Spot sizes of our $60 \pm 5$ µm MDCS excitation pulses are overlaid on the sample. (e) By spectrally resolving the signal phase, and taking the average phase shift between the dashed lines in (c), we obtain the detuning dependence of the AC-Stark effect across an inhomogeneous exciton distribution, for several pump fluences (in µJ/cm$^2$).}
\end{figure}

Figure \ref{fig:2}(a) shows the signal amplitude as a function of $t_1$ and $\hbar\omega_3$ (the energy of the emitted photons), in the absence of any pump pulse, with the sample temperature at 4 K. Two peaks at $\hbar\omega_3= 2.07$ eV and $\hbar\omega_3= 2.04$ eV are evident and attributed to the A-exciton and trion, respectively \cite{Paur_2019,plechinger2016trion,kolesnichenko2020disentangling,muir2022interactions}. 
The exciton peak decays as a function of $t_1$ due to decoherence of the ${\ket{g}\bra{X_{K^\prime}}}$ coherence, with the decoherence time, $T_2$, determined to be $360 \pm 10$ fs. The 2D spectrum, obtained by Fourier transforming with respect to $t_1$,
is shown in Appendix \ref{appen_MDCS} and is similar to previously reported 2D spectra 
\cite{moody2015intrinsic,hao2016direct,hao2016coherent,conway2022direct,muir2022interactions,hao2017neutral,jakubczyk2016radiatively,guo2020lineshape}.  

The changes to the amplitude and phase of the signal induced by a $\sigma^+$ polarized pump pulse, with fluence of 751.7 µJ/cm$^2$ and detuned by 200 meV, can be seen in Fig.\ \ref{fig:2}(b) and (c), respectively. For $t_1$ delays $<$ 250 fs, the pump arrives before $\textbf{\textit{k}}_1$ and has no influence on the signal amplitude or phase. For $t_1$ delays $>$ 400 fs, the pump arrives after $\textbf{\textit{k}}_1$, and before $\textbf{\textit{k}}_2$ and $\textbf{\textit{k}}_3$, such that the system is in a ${\ket{g}\bra{X_{K^\prime}}}$ coherence  when the pump arrives (Fig.\ \ref{fig:1}(c)). The pump induces both a drop in signal amplitude (evident for $t_1>250$ fs) and a phase shift that increase from 0 to $\pi$ between $t_1=250$~fs and $t_1=400$~fs.

It is evident in Fig.\ \ref{fig:2}(c) that the phase shift arising from the AC-Stark shift varies as a function of $\hbar\omega_3$. Figure \ref{fig:2}(e) plots the average phase shift between 400 and 500~fs for several pump fluences and clearly shows the significant decrease in phase shift as $\hbar\omega_3$ increases. The distribution of $\hbar\omega_3$ values comes from the inhomogeneous broadening of the exciton energy across the sample, as shown in Fig.\ \ref{fig:2}(d), and leads to an increase in the magnitude of the pump detuning as $\hbar\omega_3$ increases.  The $\hbar/omega_3$ dependence of the phase shift is thus consistent with the expectation that the magnitude of the AC-Stark shift scales inversely with pump detuning \cite{kim2014ultrafast,sie2015valley}. 

\subsection{\label{sec:B}Phase dynamics and total phase shift}

 \begin{figure*}
    \includegraphics[width=\textwidth]{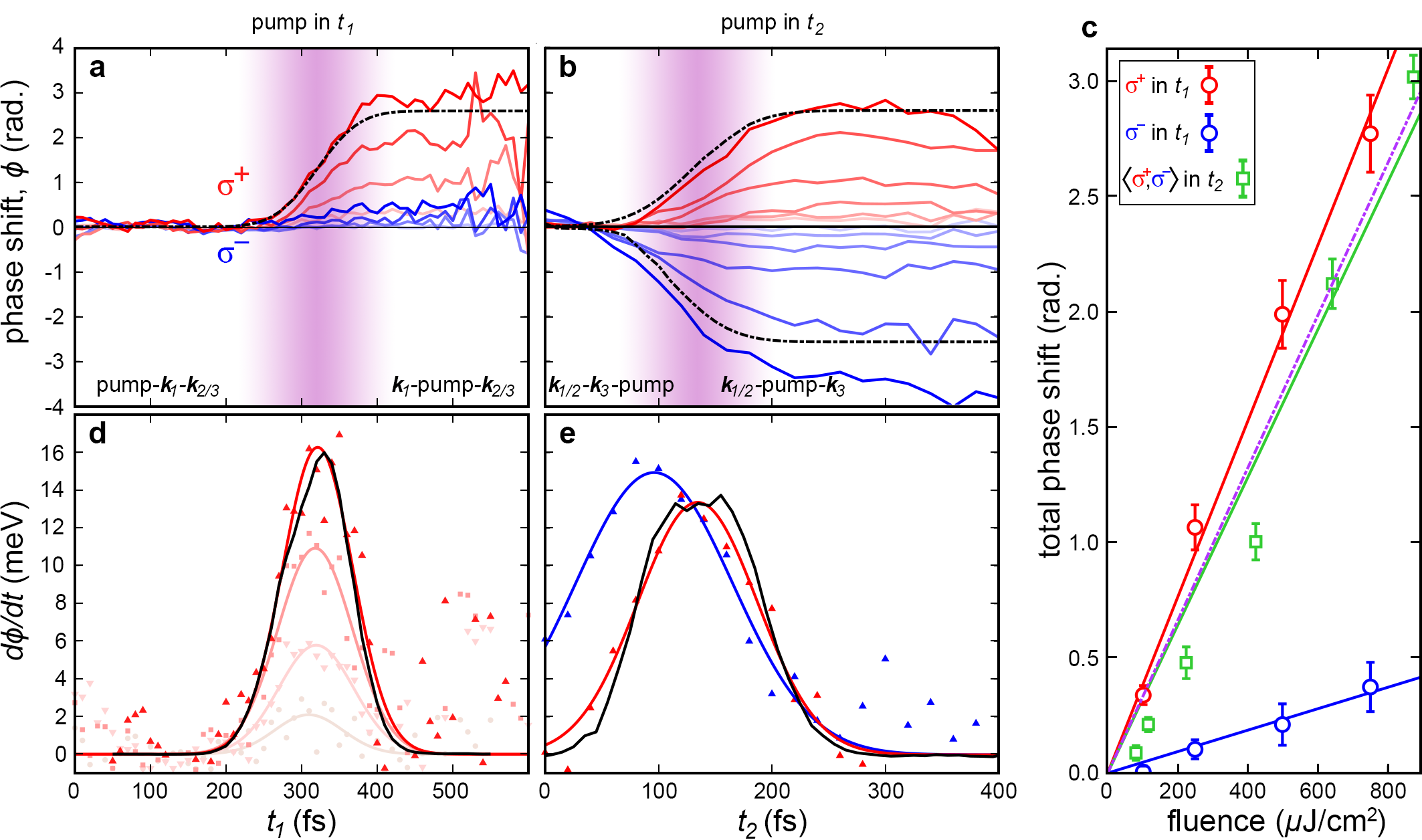}
    \caption{\label{fig:3}(a) and (b) show the phase dynamics at the exciton energy with a $\sigma^+$ (red) and $\sigma^-$ (blue) circularly polarized pump arriving during $t_1$ and $t_2$, respectively, for different fluences. Fluences in (a): 101.8, 250.0, 500.0, 751.7 µJ/cm$^2$. Fluences in (b): 81.0, 116.3, 224.7, 423.8, 643.0, 875.0 µJ/cm$^2$. Darker coloring corresponds to higher fluence. The purple shaded regions indicates the arrival of the pump pulse and the black dashed lines show the results from simulations, as described in the main text. (c) The total accumulated phase shift as a function of fluence was determined by averaging the shift at the exciton energy over a 100 fs period after the pump. For the pump in $t_2$ the magnitude of the phase shift for the $\sigma^+$ and $\sigma^-$ polarized pump pulse was averaged. The errorbars are $\pm$1 standard deviation. The lines show the linear fit to the magnitude of the phase shift as a function of pump fluence. The purple dashed line shows the difference between the fit curves for the $\sigma^+$ (AC-Stark effect) and $\sigma^-$ (Bloch-Siegert effect) polarized pump in $t_1$. This closely resembles data and fit for the pump in $t_2$. Differentiating the phase shift $\phi$ with respect to delay yields the instantaneous energy shift shown in (d) and (e) as a function of $t_1$ and $t_2$, respectively. The red and blue curves are Gaussian fits to the data. The black line shows the cross-correlation of the pump with one of the MDCS beams (on an arbitrary amplitude scale) and matches the dynamics of spectral shifts observed.}
\end{figure*}

The evolution of the measured phase difference as a function of $t_1$, obtained by taking a slice at the exciton energy ($\hbar\omega_3=2.072 \pm 0.005$ eV), is shown in Fig.\ \ref{fig:3}(a) for different pump fluences and polarizations (see Supplementary Material available at **** for the spectrally-resolved data). The induced phase shift increases as the $\textbf{\textit{k}}_1$ pulse is scanned through the pump pulse, up to a total accumulated phase shift of $\pi$ for the highest fluence. The magnitude of this phase shift varies linearly with intensity, as expected for the AC-Stark shift and shown in Fig.\ \ref{fig:3}(c). 
With the $\sigma^-$ polarized pump the AC-Stark shift occurs on the $X_K$ state, which should not affect the measured signal because the system is in a ${\ket{g}\bra{X_{K^\prime}}}$ coherence when the pump arrives and is thus insensitive to instantaneous changes in the other valley. The Bloch-Siegert effect will, however, cause a blue-shift of the $X_{K^\prime}$ state \cite{sie2017blochsiegert,slobodeniuk2022semiconductor,slobodeniuk2022giant}, and is the source of the phase shift observed in this case. This phase shift is much smaller than in the case of the $\sigma^+$ polarized pump, but still linear with fluence, as shown in Fig.\ \ref{fig:3}(c).

The magnitude of the AC-Stark shift ($\hbar\omega_{ACS}$) and Bloch-Siegert shift ($\hbar\omega_{BS}$) are given by
\begin{equation}
    \hbar\omega_{ACS}=\frac{\abs{\mu_{K^\prime}}^2E^2}{4(\hbar\omega_{X}-\hbar\omega_{p})},
    \label{eqACS}
\end{equation}
\begin{equation}
    \hbar\omega_{BS}=\frac{\abs{\mu_K}^2E^2}{4(\hbar\omega_{X}+\hbar\omega_{p})},
    \label{eqBS}
\end{equation}
where µ is the dipole matrix element for the $K$ and $K^\prime$ excitons and $E$ is the electric field amplitude \cite{sie2017blochsiegert}. With a red-detuned pump energy of $\hbar\omega_{p}=$ 1.87 eV and exciton energy $\hbar\omega_X=$ 2.07 eV, the expected ratio between the AC-Stark shift and Bloch-Siegert shift is $(\omega_X+\omega_{p})/(\omega_X-\omega_{p})=$ 19.7. The measured ratio of slopes in Fig.\ \ref{fig:3}(c) is $8 \pm 3$. This suggest the measured Bloch-Siegert shift is approximately a factor of two larger than would be expected based on the measured AC-Stark shift. This discrepancy cannot be explained by the polarization purity of the pump or MDCS pulses (as discussed in detail in the Supplementary Material), but may arise from Coulomb interactions and the presence of higher energy bands in WS$_2$, which are excluded from the two-level framework used to derive Eq. 1 and 2, and are expected to affect the AC-Stark and Bloch-Siegert shifts differently \cite{slobodeniuk2022semiconductor,slobodeniuk2022giant,kobayashi2023floquet}.

To observe the effect of the pump on the inter-valley coherence, the pump was shifted to arrive 150 fs after $\textbf{\textit{k}}_1$ and $\textbf{\textit{k}}_2$, which arrived simultaneously. The phase of the signal at the exciton energy is plotted in  Fig.\ \ref{fig:3}(b) as a function of $t_2$. Here, for $t_2<100$ fs the pump arrives after $\textbf{\textit{k}}_3$ and actually acts on the $\ket{X_K}\bra{g}$ coherence. For $t_2>200$ fs the pump is acting on the inter-valley coherence. Figure \ref{fig:3}(b) shows that phase shifts with approximately equal magnitude, but opposite sign, were observed for the two pump polarizations. The evolution of the inter-valley coherence $\ket{X_K}\bra{X_{K^\prime}}$ is given by $e^{i(\omega_{X_K}-\omega_{X_{K^\prime}})t_2}$ and shifts to either the $X_K$ or $X_{K^\prime}$ energy will lead to a phase shift. With the highest pump fluence of 875 µJ/cm$^2$ a $\pi$ phase shift of the inter-valley exciton coherence was realized.

The total phase shift, obtained by averaging the phase shift between $t_2=220$~fs and $t_2=300$~fs, is plotted as a function of pump fluence in Fig.\ \ref{fig:3}(c) (green line). A linear trend is again observed, however, the slope is slightly smaller than for the case of the AC-Stark shift on the $\ket{g}\bra{X_{K^\prime}}$ coherence. In fact, the slope of the total phase shift when the pump is in $t_2$ and acting on the $\ket{X_K}\bra{X_{K^\prime}}$ coherence is equivalent to the difference between the isolated slopes of the AC-Stark effect and Bloch-Siegert effects (dashed line in Fig.\ \ref{fig:3}(c)), acquired when the pump was in $t_1$. This arises because $X_{K^\prime}$, and $X_K$ are both blue-shifted (by the AC-Stark shift and Bloch-Siegert shift, respectivley), and thus the energy difference between them (and hence accumulated phase shift) is given by the difference between the two effects. That is, the two effects act to counter one another when rotating the phase of the $\ket{X_K}\bra{X_{K^\prime}}$ coherence. 

The shaded regions in  Fig.\ \ref{fig:3}(a) and (b) indicate the delay range where the pump overlaps with the pulse being scanned ($\textbf{\textit{k}}_1$ and $\textbf{\textit{k}}_3$ respectively). In these regions the dynamics of the phase accumulation can be observed.
The rate of change of the phase is determined by the induced energy shifts, i.e.\ the coherence evolves as $e^{i\Delta t_n}$. The time dependent energy shift can thus be determined by differentiating the measured phase shift: $\Delta(t_n)=d\phi/dt_n$. This is shown in Fig.\ \ref{fig:3}(d) for a $\sigma^+$ polarized pump in $t_1$.  A Gaussian fit to the instantaneous AC-Stark shift (red lines) closely resembles the temporal profile of the measured cross-correlation between the pump and the MDCS excitation beams (black line). This indicates that within the temporal resolution afforded by the 31 fs $\textbf{\textit{k}}_1$ pulse, the AC-Stark shift follows the  intensity of the pump pulse envelope. The peak AC-Stark shift measured is $>$16 meV for the highest pump fluence of 751.7 µJ/cm$^2$ \footnote{The measured evolution of the {Stark} shift arises from a convolution of the $\textbf{\textit{k}}_1$ pulse and the pump-induced shift, hence the actual peak shift will be slightly higher than the one obtained directly from the measurement}. Similarly, with the $\sigma^+$ polarized pump in $t_2$ the measured time dependent AC-Stark shift follows the cross-correlation between the pump and the MDCS beam (Fig.\ \ref{fig:3}(e)). With the $\sigma^-$ polarized pump, the peak is broader and shifted to shorter $t_2$ times because in this case,  for $t_2<100$ fs, the system is in a $\ket{X_K}\bra{g}$ coherence when the pump arrives, and $X_K$ is affected by the AC-Stark shift.

The dynamics of the AC-Stark and Bloch-Siegert shifts are modelled by incorporating these shifts as phenomenological perturbations to the state energies in simulations of the MDCS experiments (see Appendix \ref{appen_SIMS} for details). There were no free parameters in the model (all were determined directly from experiment) and it was assumed that the AC-Stark and Bloch-Siegert shifts followed exactly the instantaneous intensity of the 100 fs pump pulse. The results are shown as dashed lines in Fig.\ \ref{fig:3}(a) and (b). The close agreement between experiment and simulation further suggests that all details of the shifts are captured in the model and that the phase evolves smoothly as the energy shifts follow the instantaneous intensity of the pump pulse.

\subsection{\label{sec:C}Power broadening and loss of macroscopic coherence}

\begin{figure}
    \includegraphics{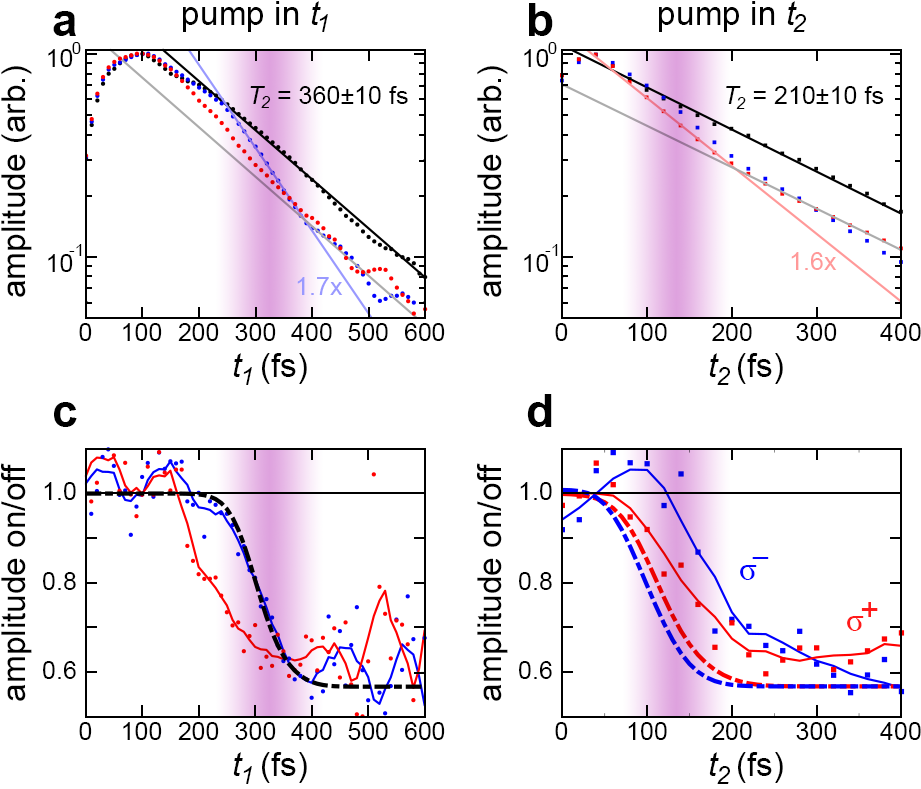}
    \caption{\label{fig:4}(a) Signal amplitude (log scale) integrated around the exciton energy as a function of $t_1$. The unpumped data (black points) was fit with a linear slope (black line) to yield a decoherence time $T_2 = 360 \pm 10$ fs. The red and blue data points show the amplitude in the case of a the 751.7 µJ/cm$^2$ fluence pump pulse $\sigma^+$ and $\sigma^-$ polarized, respectively. Through the pulse duration (shaded region) the amplitude decreases due to power broadening of the exciton linewidth. For the period $270<t_1>380$ a linear fit (blue line) gave a slope with decay rate 1.7 times faster than the unpumped decay. After the finite duration of the pump, the decay returns to the unpumped rate as shown by the grey trend line from a linear fit to the pumped data between $380<t_1>600$. (b) Signal amplitude (log scale) integrated around the exciton energy as a function of $t_2$. A fit to the unpumped data (black line) yields a decoherence time $T_2$ of $210 \pm  10$ fs for the inter-valley coherence. In the presence of the pump pulse with fluence 423.8 µJ/cm$^2$ (red and blue data points), the decay rate increases by a factor of 1.6 over the  pump duration. (c) and (d) plot the signal amplitudes of the pumped data divided by the unpumped reference as a function of $t_1$ and $t_2$, respectively. It is clear that the signal amplitude reduces due to the pump, and is flat beyond the pump duration. The lines follow the smoothed data points. The dashed lines show the results of phenomenological simulations, as discussed in the main text.}
\end{figure}

Despite the smooth evolution of the phase, the decrease in the signal amplitude seen in Fig.\ \ref{fig:2}(b) suggests the pump causes a partial loss of coherence, and thus that the interaction of the red-detuned pump is not entirely adiabatic. Figure \ref{fig:4}(a) and (b) show the normalised signal amplitude, integrated around the exciton energy, as a function of $t_1$ and $t_2$, respectively. In both cases, a decrease in the signal amplitude, is evident upon arrival of the pump pulse, which indicates additional decoherence of the $\ket{g}\bra{X_{K^\prime}}$ and $\ket{X_K}\bra{X_{K^\prime}}$ coherences, regardless of the pump polarization. After the pump pulse, the decay rate once again matches the unpumped data, as indicated by the slope of the data and fits in Fig.\ \ref{fig:4}(a) and (b). The return of the decay rate after the pump pulse is even more clearly demonstrated in  Fig.\ \ref{fig:4}(c) and (d), which show the ratio of the pumped and unpumped amplitudes. It is clear that for $t_2>200$ fs, the curves are flat, indicating that the decay rates with and without the pump are the same \footnote{For the pump in $t_1$, the curve for $t_1>400$ fs is flat for the simple reason that coherence time before the pump is increasing}. This rules out the possibility that two-photon absorption of the pump-pulse is creating additional carriers that scatter and cause additional decoherence, as they would persist and affect the decoherence time well beyond the pulse overlap range. Thus, we conclude that the enhanced decoherence rate is consistent with a field driven effect persisting only over the finite pump duration. In the measurements of Ye et al.\ \cite{ye2017optical} they observe a decrease of the degree of linear polarization, which could reflect a loss of coherence. However, in that case they attribute those losses to the fact that the response is integrated in time, and includes emission before, during, and after the phase is rotated by the pump. They also have some contribution from the Gaussian distribution of pump intensity, which will cause different phase shifts. In our measurements, both of these effects are eliminated: the measured signal is proportional only to the coherence at the time the 30 fs $\textbf{\textit{k}}_3$ pulse interacts with the sample, hence there is minimal temporal averaging. The spatial averaging over a distribution of pump intensities was minimised by ensuring that the pump spot size was larger than the MDCS spots and because the MDCS signal scales with the product of the electric field of each pulse, the effective area for the signal generation has a diameter a factor of $\sqrt{2}$ smaller than the individual beams. It is therefore clear that in our measurements there is some additional non-trivial decoherence when the pump pulse is present.

The magnitude of the extra coherence loss is linearly proportional to the pump fluence (see Supplementary Material) and is quantitatively consistent with power broadening.
For the case of a constant wave or very long pump pulse, the power broadened linewidth is given by \cite{foot2004atomic}:
\begin{equation}
    \Gamma = \Gamma_0\left(1+\frac{2\Omega^2}{\Gamma_0^2}\right)^{1/2}
    \label{powerbroadening}
\end{equation}
where $\Omega=\abs{\mu}E$ is the resonant Rabi splitting \cite{xu2008single} \footnote{as opposed to the generalized Rabi splitting, which takes into account the detuning}, and $\Gamma_0$ is the equilibrium homogeneous linewidth. The ratio $\frac{\Omega}{\Gamma_0}$ indicates that power broadening is effectively dependent on the number of Rabi oscillations that occur within the decoherence time. In the case of short pulses, where the pulse duration is less than the decoherence time, quantifying the power broadening can be more challenging \cite{vitanov2001power}. The amount of broadening is still expected to depend on the Rabi frequency, and hence vary with the intensity across the pulse envelope \cite{vitanov2001power}. However, the decoherence time is no longer the shortest relevant timescale. To place an upper limit on the peak power broadening, we calculated the Rabi frequency from Eq.\ \ref{eqACS}, using the peak AC-Stark shift measured (16 meV), and in place of $\Gamma_0$ we use the inverse of the pump pulse duration: $=\frac{1}{100 fs}=41$ meV. This gives a value of 6.9 meV for the power broadened linewidth, compared to the equilibrium value of 1.8 meV.

An experimental estimate of the average power broadening throughout the pump pulse is obtained from the data in Fig.\ \ref{fig:4}(a) and (b). A linear fit to the data (on the semi-log axes) over the duration of the pulse yields an average decay rate that is 1.6-1.7 times faster than the unpumped decay. This suggests an average power broadened linewidth of $\sim$3 meV.

The power broadening was added to the simulations as a phenomenological contribution to the homogeneous linewidth, which follows the time dependent intensity of the pump pulse. This is an oversimplification, as the process for power broadening is not expected to be instantaneous. Nonetheless,  using a peak broadening of 3.5 meV (i.e.\ at the peak of the pump pulse the power broadened linewidth is 5.3 eV) the simulated signal amplitude matches the dynamics of the measured amplitude for the $\sigma^-$ case exceptionally well, as shown in Fig.\ \ref{fig:4}(c). This value for the peak broadening is consistent with the upper limit calculated above and the average linewidth determined by fitting a single decay to the data in the pump overlap region. For the other measurements, the dynamics and/or magnitude of the decrease in amplitude differs somewhat from the simulations using the same parameters. These differences highlight the fact that the dynamics of power broadening for short pulses, and with the valley selectivity of monolayer TMDCs, are more complicated than the simple picture considered. However, with our phenomenological model, and quantitative comparisons, we have demonstrated that power broadening is present and the associated loss of coherence presents a limitation when using the AC-Stark and Bloch-Siegert effects to control bandstructures and the phase of quantum coherences.

\section{\label{sec:IV}Conclusions}

By directly creating and measuring the phase of inter-valley coherences using MDCS, we have demonstrated the ability to measure the coherent dynamics while a red-detuned pump pulse drives Floquet-Bloch states. With a fluence of 875.0 µJ/cm$^2$ a $\pi$ phase-shift occurred over the 100 fs pump pulse duration. The dominant cause of the phase shift was the valley selective AC-Stark shift, which lifts the degeneracy of the $K$ and $K^\prime$ excitons. However, the energy difference between excitons in the $K$ and $K^\prime$ valleys induced by the AC-Stark effect is reduced by the Bloch-Siegert effect, which blue-shifts the exciton in the other valley. These two effects were isolated by measuring the coherence between ground and single exciton states in the presence of the pump. 

The smooth evolution of the phase suggests the process is adiabatic, but in these ensemble measurements the amplitude of the macroscopic coherence provides a better measure. Additional, irreversible, decoherence due to power broadening in the presence of the pump field was revealed, and imposes limitations on applications requiring adiabatic band control. The extent of the decoherence scales linearly with the pulse fluence, but so too do many of the effects targeted for control. Nonetheless, with this understanding of the role of power broadening, and depending on the application, it may be possible to minimise these effects. For example, to reduce power broadening and still achieve a $\pi$ phase shift of the inter-valley coherence, the pump laser detuning could be reduced (while still ensuring no spectral overlap between laser and exciton spectra), which will increase the AC-Stark shift and decrease the Bloch-Siegert shift for a given pulse fluence.

To reduce power broadening more generally, it will be necessary to understand and quantify the effect for short pulses and build models beyond the simplicity of a two-level system. Further potential lies in the idea that such pulses can be shaped to induce spectral narrowing \cite{boradjiev2013power}, which may open new options for control. The experiments described here, and their ability to measure the amplitude and phase of coherences, provide an ideal means to explore and test these possibilities.


\begin{acknowledgments}
This work was funded by the Australian Research Council Centre of Excellence in Future Low-Energy Electronics Technologies (CE170100039). T.H.Y.V., M.S.F., and M.T.E.\ acknowledge funding support from DP200101345. K.W. and T.T. acknowledge support from JSPS KAKENHI (Grant Numbers 19H05790, 20H00354 and 21H05233). We thank Jared Cole for fruitful discussions.
\end{acknowledgments}


\appendix

\section{Sample preparation}
Hexagonal boron nitride (hBN) crystals were mechanically exfoliated onto a 285 nm silicon dioxide substrate. The monolayer WS$_2$ (HQ Graphene) was prepared on a polydimethylsiloxane (PDMS) substrate using the same method. Next, the monolayer WS$_2$ was transferred onto the hBN using a dry transfer method. For encapsulation, a thin hBN flake ($<$10 nm) was picked up by poly carbonate (PC) film. The hBN thin flake and the film were then dropped down to cover the WS$_2$ monolayer. Finally, the sample was cleaned in chloroform for 5 minutes to dissolve the polymer and annealed in ultra-high vacuum for 8 hours at 200 degrees Celsius. 

\begin{figure}
    \includegraphics{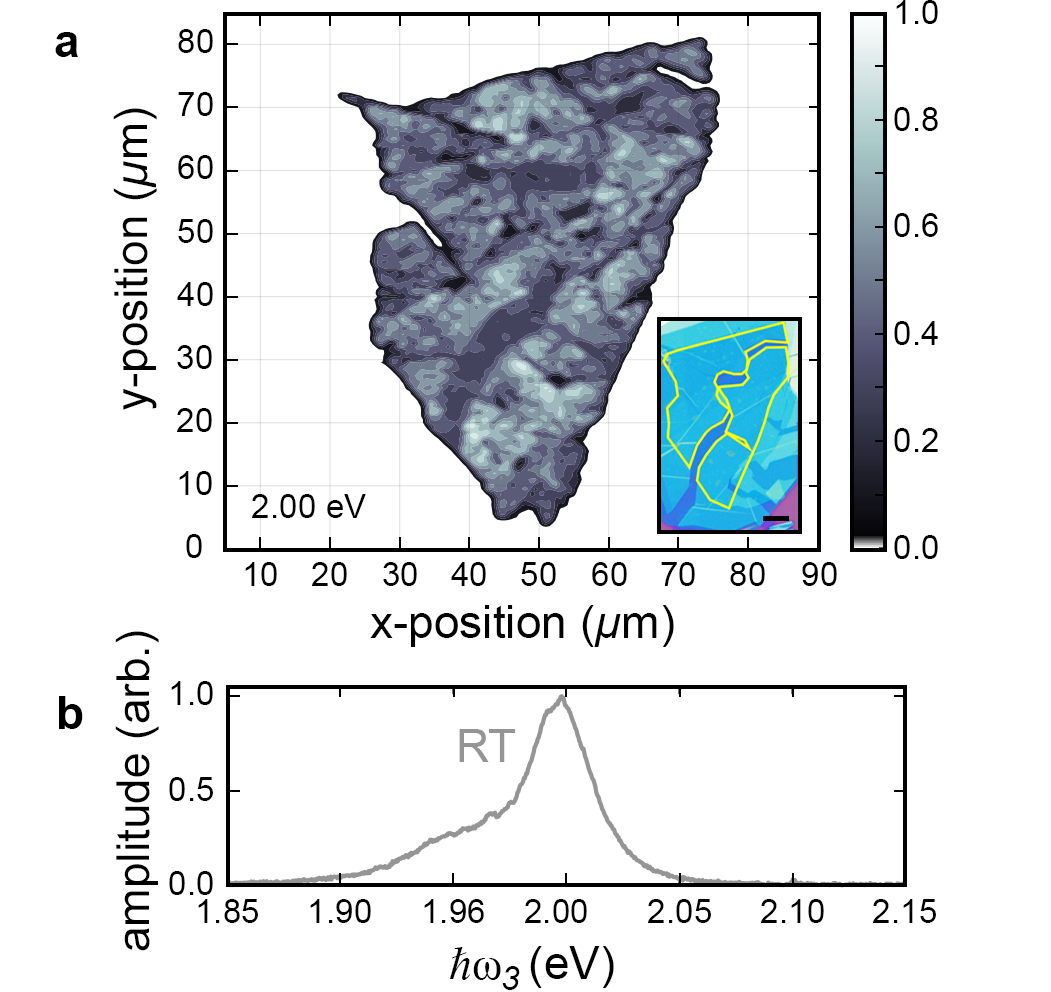}
    \caption{\label{fig:5}(a) Spatially resolved PL intensity at 2.00 eV across the sample. Micro-cracks, bubbling hBN encapsulation, and other defects lead to inhomogeneity, which can be seen as variations in PL intensity across the sample surface. The $60 \pm 5$ µm spot sizes used in our MDCS measurements sample most of this inhomogeneity. Inset: Optical image of the sample. The top hBN broke during transfer and the encapsulated monolayer region is marked in yellow. The scale bar is 10 µm. (b) Room temperate PL averaged over the entire sample.}
\end{figure}

Room temperature spatially resolved photoluminescence (PL) was used for initial characterization of the hBN encapsulated monolayer WS$_2$ flake used throughout this work, shown in Fig.\ \ref{fig:5}(a). In our setup, a cw laser diode (Thorlabs, L405P20) tuned to $\sim$410 nm is coupled to a single-mode fiber (Thorlabs, P1-460B-FC-1), the output of which is reflected by a dichroic mirror (Edmund Optics, longpass 492 nm) into a 60$\times$ objective lens (Olympus, NA=0.8), and focused onto the sample. PL emission is collected by the same objective lens, transmitted through the dichroic, and sent into a free-space fiber coupler (Thorlabs, F810SMA-635) which couples the light into a miltimode fiber (Avantes, 200 µm core diameter). The fiber is connected directly to a spectrometer (Avantes, Avaspec-2048). Combined with an X-Y motorised translation stage (Applied Scientific Instrumentation, MS-2000), we raster across the sample taking spectra at each point. The resultant room temperature PL map is shown in Fig.\ \ref{fig:5}(a) (and main text Fig.\ \ref{fig:2}(d)), where we have mapped the normalized PL intensity at 2.00 eV. Inhomogeneity in the sample is evident, and can be due to defects, cracking, and bubbling from the hBN encapsulation. The room temperature (RT) sample averaged PL spectrum is shown in Fig.\ \ref{fig:5}(b). 

\section{\label{appen_MDCS}Multidimensional coherent spectroscopy}
\begin{figure*}
    \includegraphics[width=\textwidth]{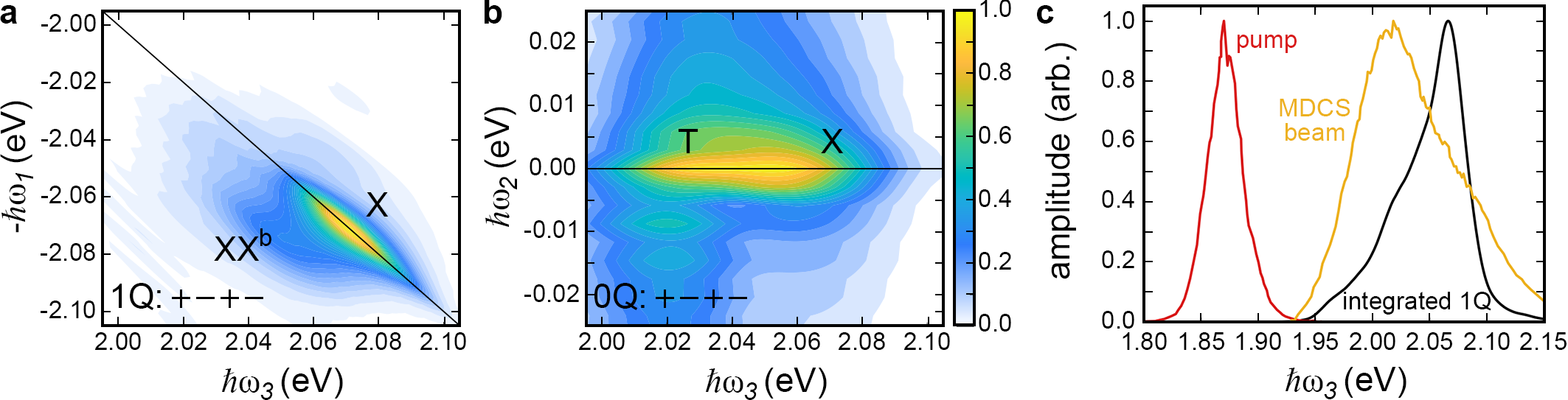}
    \caption{\label{fig:6}(a) Unpumped 1Q 2D amplitude spectrum with cross-polarized excitation pulses, showing a bright exciton peak (X) and the biexciton peak (XX$^b$), red-shifted in emission energy ($\hbar\omega_3$). The sample temperature was 4 K. (b) Unpumped 0Q 2D amplitude spectrum features peaks due to an inter-valley exciton coherence (X) and trion coherence (T). All observations are consistent with previous MDCS measurements on monolayer TMDCs \cite{conway2022direct,moody2015intrinsic,hao2016coherent,muir2022interactions}. (c) Integrated 1Q sample response along $-\hbar\omega_1$ from (a) (black line) is plotted along with the MDCS excitation spectrum (yellow) and the pump spectrum (red).}
\end{figure*}

Our MDCS pulses were generated by a Yb:KGW laser amplifier (Light Conversion Pharos), which pumps a non-colinear optical parametric amplifier (Light Conversion Orpheus-N-3H) tuned to output pulses with a spectrum centred near 2.00 eV (Fig.\ \ref{fig:6}(c)). These pulses from the NOPA were incident onto a 2D grating and the four first-order diffracted beams selected as our three MDCS excitation pulses and local oscillator (LO). Independent spectral amplitude and phase control of each beam was enabled by a pulse shaper based on a spatial light modulator (Santec LCOS-SLM-100) \cite{tollerud2017coherent,turner2011invited}. The pulse shaper also facilitates correction of temporal chirp introduced by the optics and beam propagation to the sample, by using the multi-photon intra-pulse interference phase scan technique \cite{xu2006quantitative}. This ultimately led to pulses $\sim$30fs in duration. A high degree of phase stability was passively maintained by ensuring the four beams (or pairs of beams) were incident on common optics \cite{tollerud2017coherent}.

All beams were focused through a 75 mm plano-convex lens onto the sample, which was in a cryostat (Montana Instruments—Cryostation) at 4 K and low pressure. The four beams were spatially overlapped with a spot size of $60 \pm 5$ µm FWHM. Beam fluences at the sample position were kept below 2 µJ/cm$^{2}$ per pulse to ensure the contributions of signals beyond the $\chi^{(3)}$ regime were insignificant.
At the sample surface, three light-matter interactions occur from excitation pulses with wavevectors $\textbf{\textit{k}}_1$, $\textbf{\textit{k}}_2$, and $\textbf{\textit{k}}_3$ to produce a four-wave mixing (FWM) signal emitted in the background free direction given by conservation of momentum. We measure the signal emitted in the direction $\textbf{\textit{k}}_{sig}=-\textbf{\textit{k}}_1+\textbf{\textit{k}}_2+\textbf{\textit{k}}_3$, which is spatially overlapped with the LO, on the fourth corner of the box, for heterodyne detection. The signal and LO were subsequently sent into a spectrometer with a CCD detector to obtain a spectral interferogram, enabling the spectrally-resolved signal amplitude and phase to be determined.

For the 1Q rephasing measurements discussed in the main text, the signal was measured as a function of $t_1$ by scanning the arrival time of $\textbf{\textit{k}}_1$, with $t_2=0$. For the 0Q measurements, $t_2$ was scanned by delaying $\textit{\textbf{k}}_3$, with $t_1=0$. $\hbar\omega_3$ is the Fourier transform of $t_3$, and is measured directly by the spectrometer. The polarization of the pulses was set using half-wave plates in each beam to rotate the linear polarization of each pulse independently. A quarter-wave plate common to all beams was used to convert the orthogonal linear polarizations to be circularly polarized, either $\sigma^+$ or $\sigma^-$. For the measurements described here, the $\textbf{\textit{k}}_1$ and $\textbf{\textit{k}}_3$ pulses were set to be $\sigma^+$, while $\textbf{\textit{k}}_2$ and $\textbf{\textit{k}}_{LO}$ pulses were set to be $\sigma^-$ polarized.

An additional advantage of measuring both the amplitude and phase of the signal as a function of the delays is that there is a unique solution to the Fourier transform with respect to these delays. This enables the generation of multi-dimensional spectra, as shown in Fig.\ \ref{fig:6}(a) and (b) for the unpumped 1Q and 0Q rephasing data, respectively. 
The 1Q rephasing 2D spectrum features a diagonal peak (for which $-\hbar\omega_1=\hbar\omega_3$), corresponding to excitation and emission at the exciton energy (X), and a cross peak red-shifted from the exciton emission energy, which we attribute to the neutral biexciton (XX$^b$) \cite{conway2022direct}. 
In our MDCS measurements, the $60 \pm 5$ µm spot size of the excitation pulses averages over the sample inhomogeneity, resulting in broadening along the main diagonal (black line in Fig.\ \ref{fig:6}(a)). Despite this, the width of the 2D peak across the diagonal is much narrower and represents the homogeneous linewidth. In the unpumped 0Q 2D spectrum the peaks sit along the horizontal line at $\hbar\omega_2$=0 meaning the coherence produced between inter-valley excitons are from energetically degenerate states. We also observe an inter-valley trion coherence, labelled T. 
These 2D spectra are consistent with previous MDCS measurements on monolayer TMDCs \cite{conway2022direct,moody2015intrinsic,hao2016coherent,muir2022interactions}. 

\section{Pump characteristics}

Pulses were generated by the same Yb:KGW laser amplifier as above, which also pumps a second NOPA (Light Conversion Orpheus-N-2H) tuned to output pulses with a spectrum centred 200 meV below the exciton transition at 1.87 eV (Fig.\ \ref{fig:6}(c)), and 100 fs in duration. These pulses travel down a computer controlled motorized delay stage with a retroreflector at its end to control the pump timing. The pump then passes through a mechanical shutter, and finally a liquid crystal retarder (Thorlabs LCC1111-D) and controller (Thorlabs LCC25). The axis of the liquid crystal retarder was set to be 45$\degree$ to the pump beam polarization, and the voltage set such that there phase shift of zero (polarization unchanged) or $\lambda/2$, in which case it acts as a half-wave plate and rotates the polarization 90$\degree$. The pump was then directed via a pick-off mirror to be parallel to, but slightly offset from the MDCS beams, so that it passes through the same quarter-wave plate and focusing lens onto the sample, and overlaps with the MDCS beams. The spot size of the pump was $\sim90$ µm. An optical isolator was constructed at the sample position to set the voltage of the liquid crystal retarder such that two polarization states of the pump could be defined; co- and cross-polarized to $\textit{\textbf{k}}_1$. 
For each delay position of the 1Q and 0Q scans, data was acquired for both pump polarizations, and without the pump (i.e. with the shutter closed) to optimise the ability to identify changes to the signal resulting from the $\sigma^+$ and $\sigma^-$ polarized pump. 

\section{\label{appen_SIMS}Simulation details}

To simulate the MDCS experiments we used an approach based on perturbatively solving the Liouville formulation of the Schrodinger equation ($i\hbar\Dot{\rho} = [H,\rho]$) for a three-level``V-type" system, similar to that used in Ref.\ \cite{JAD_JCP_2011,JAD_PRB_2007}. The Hamiltonian is given by:
\begin{equation*}
H=\begin{bmatrix} \hbar\omega_1 & \mathbf{E\cdot\mu_{12}} & \mathbf{E\cdot\mu_{13}} \\ \mathbf{E\cdot\mu_{21}} & \hbar\omega_2 & 0\\\mathbf{E\cdot\mu_{31}} & 0 & \hbar\omega_3 \end{bmatrix}  
\end{equation*} 
where $\hbar\omega_n$ is the energy of state n. For the case of monolayer WS$_2$, state 1 is the ground state, and state 2 and 3 are the degenerate exciton states, with energy set to be 2.06~eV. $\mu_{ij}$ is the transition dipole moment for the transition between state \textit{i} and \textit{j}, and $\mathbf{E}$ describes the electric field of the three laser laser pulses: 
\begin{equation*}
    E=\sum_n E_n e^{i(\omega_L(t-\tau_{n})-\mathbf{k_n}.\mathbf{r})}e^{-\frac{(t-\tau_n)^2}{2c^2}}  + c.c.
\end{equation*}
where $E_n$ represents the peak amplitude of each pulse, $\tau_n$ is the arrival time of pulse  n, $\omega_L$ is the centre laser frequency (2.06~eV), and c is the standard deviation of the Gaussian pulse envelope (12.7~fs, corresponding to FWHM = 30~fs).
To account for the selection rules each pulse is multiplied by a vector 
$\begin{bmatrix}
    1  0
\end{bmatrix}$
or $\begin{bmatrix}
    0  1
\end{bmatrix}$ for a $\sigma^+$ or $\sigma^-$ polarized pulse, respectively. The transition dipole moment for state 2 and 3, corresponding to $K$ and $K^\prime$ excitons are multiplied by $\begin{bmatrix}
    0  1
\end{bmatrix}$
or $\begin{bmatrix}
    1  0
\end{bmatrix}$
 respectively. This ensures that a $\sigma^+$ ($\sigma^-$) polarized pulse can only drive a transition in the $K^\prime$ ($K$) valley.

Relaxation is introduced phenomenologically such that 
\begin{equation*}
    i\hbar\Dot{\rho}=[H,\rho]-\begin{bmatrix} \Gamma_{11}\rho_{11} & \Gamma_{12}\rho_{12} & \Gamma_{13}\rho_{13} \\ \Gamma_{21}\rho_{21} & \Gamma_{22}\rho_{22} & \Gamma_{23}\rho_{23}\\ \Gamma_{31}\rho_{31} & \Gamma_{32}\rho_{32} & \Gamma_{33}\rho_{33}
        
    \end{bmatrix}
\end{equation*}
    
where $\Gamma_{ij}$ is the decoherence rate of the $\ket{i}\bra{j}$ coherent superposition, (set to be $1/350$ fs for all coherences to match the experimental data), and $\Gamma_{ii}$ is the population relaxation rate for state $i$, which was set to be $1/3000$ fs, but for these simulations, which do not measure any population dynamics, this value is unimportant.

Treating the electric field as a perturbation and expanding the equations perturbatively leads to a set of interdependent differential equations. By separating these in terms of the wavevector component, and keeping just the contributions that will lead to a third order polarization with wavevector $-\textbf{\textit{k}}_1+\textbf{\textit{k}}_2+\textbf{\textit{k}}_3$, this reduces to 16 equations \cite{JAD_PRB_2007}. These equations were solved numerically to give the signal electric field as a function of $t_3$ for a given set of pulse delays. The data was Fourier transformed with respect to $t_3$ to give the $\hbar\omega_3$  axis. To obtain the 1Q and 0Q 2D datasets, the arrival time of pulse 1 and pulse 3 were varied, respectively.

\subsection{Incorporating the spectral shift and power broadening}
The effects of the pump pulse were introduced phenomenologically, based on the assumption that the magnitude of the shift and broadening scale directly with the instantaneous intensity of the pump laser pulse. The time dependent energy shift was calculated as 
\begin{equation}
    \Delta_{\hbar\omega}(t)=\delta_{\hbar\omega} e^{-\frac{(t-t_{pump})^2}{2c^2}}
\end{equation}
while the time dependent broadening was given by 
\begin{equation}
    \Delta_\Gamma(t)=\delta_\Gamma e^{-\frac{(t-t_{pump})^2}{2c^2}}
\end{equation}
where $t_{pump}$ is the arrival time of the pump, $c$ is the standard deviation of the Gaussian pump pulse envelope (42.3~fs, corresponding to FWHM = 100~fs), $\delta_{\hbar\omega}$ is the peak energy shift for a given fluence, and $\delta_{\Gamma}$ is the peak increase in the broadening.

\begin{figure*}
    \includegraphics[width=\textwidth]{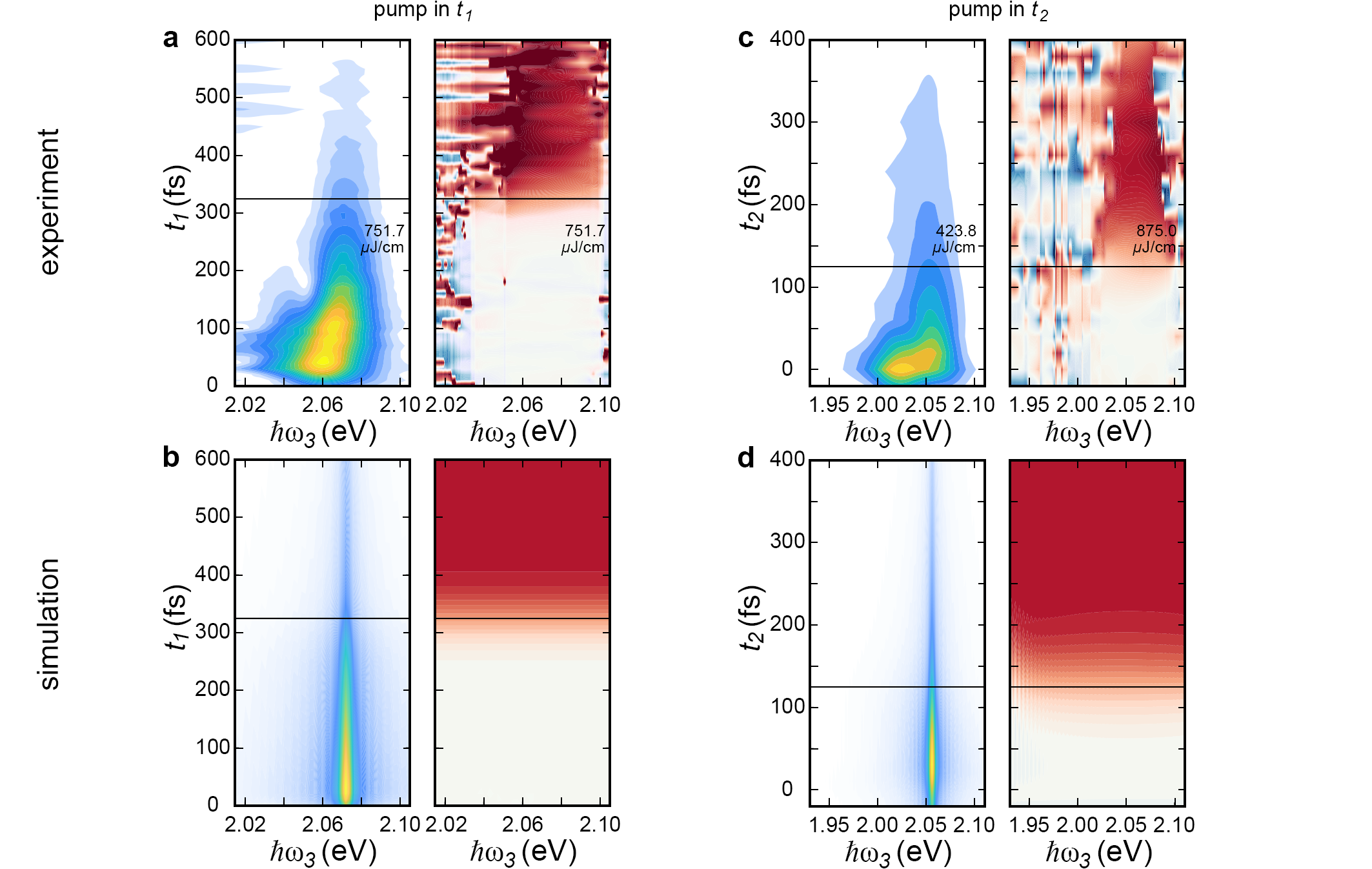}
    \caption{\label{fig:S6}(a) and (b) are the experimental data and simulation data, respectively, for a $\sigma^+$ polarized pump in $t_1$ with a fluence of 751.7 µJ/cm$^2$. (c) and (d) are the experimental data and simulation data, respectively, for a $\sigma^+$ polarized pump in $t_2$. The amplitude spectra have a pump with a fluence of 423.8 µJ/cm$^2$, while the phase spectra are shown with a pump fluence of 875.0 µJ/cm$^2$.}
\end{figure*}

The $\Delta_{\hbar\omega}(t)$ was added to the energy of state 2 or state 3, depending on the polarization of the pump. While the $\Delta_{\Gamma}(t)$ was added to the decoherence rates, $\Gamma_{ij}$. The results of the simulations are shown in Fig.\ \ref{fig:S6}(b) and (d) and compared to the highest fluence experimental data ((a) and (c)) for the 1Q and 0Q scans with the pump in $t_1$ and $t_2$, respectively. $\Delta_{\hbar\omega}$ was set to 16~meV to match the value determined from the experiments. The value of $\Delta_{\Gamma}(t)$ was 3.5meV, which was optimised to give a drop in the simulated signal amplitude that matched the experimental drop observed for the case of the pump in $t_1$.

A key difference between the simulations and the experimental data is that inhomogeneous broadening was not included in the simulations, which means the $\hbar\omega_3$ spectrum is narrower than the experiments where there is significant inhomogeneous broadening and also the presence of a trion. This also means that the phase shift seen in the simulations does not vary as a function of $\hbar\omega_3$, but is instead uniform. The absence of noise also means the phase shift is seen even on the wings of the peak where the amplitude is low. The spectrally integrated phase and amplitude are plotted as a function of delay in Fig.\ \ref{fig:3} and Fig.\ \ref{fig:4}, respectively, and show good agreement between simulations and experiments.

\bibliography{references}

\end{document}